\documentclass[prl,twocolumn,a4paper]{revtex4}
\usepackage{epsfig}
\usepackage{amsfonts}
\usepackage{amsmath}
\usepackage{graphicx}

\newcommand{\eq}{\begin{equation}}
\newcommand{\eqx}{\end{equation}}
\newcommand{\eqn}{\begin{eqnarray}}
\newcommand{\eqnx}{\end{eqnarray}}
\newcommand{\f}[2]{\frac{#1}{#2}}

\newcommand{\al}{\alpha}

\newcommand{\om}{\omega}
\newcommand{\Dl}{\Delta}

\newcommand{\nn}{{\cal N}}

\begin{document}

\title{Gauge/gravity duality and thermalization of a
  boost-invariant perfect fluid}  

\author{Romuald A. Janik}\email{ufrjanik@if.uj.edu.pl}
\affiliation{Institute of Physics, Jagellonian University,
Reymonta 4, 30-059 Krakow, Poland.}
\author{Robi Peschanski}\email{pesch@spht.saclay.cea.fr}
\affiliation{CEA/DSM/SPhT,
Unit\'{e} de Recherche associ{\'e}e au CNRS,
CEA-Saclay, F-91191 Gif/Yvette Cedex, France.}

\begin{abstract}
We derive the equation for the quasi-normal modes corresponding to the
scalar excitation of a black hole moving away in the fifth
dimension. This geometry is the AdS/CFT dual of a boost-invariant expanding
perfect fluid in  ${\cal N}=4$ SUSY Yang-Mills theory at large
proper-time. On the gauge-theory side, the dominant solution of the equation
describes the decay back to equilibrium of a scalar excitation of the
perfect fluid. Its characteristic  proper-time can
be interpreted as a thermalization time of the perfect fluid, which is
a universal (and numerically small) constant in units of the unique scale of the problem. This may provide a new insight on
the short thermalization-time puzzle encountered in heavy-ion collision
phenomenology. A nontrivial scaling behaviour in proper-time is obtained which can be interpreted in terms of a slowly varying adiabatic approximation.  
\end{abstract}

\maketitle

\subsection{Thermalization response-time and Black Hole quasinormal modes}
In a recent paper \cite{1}, we have shown that the AdS/CFT dual of an expanding
relativistic thermal and perfect fluid in ${\cal N}=4$ SUSY Yang-Mills (SYM)
theory can be identified as a black hole (BH) moving away in the fifth
dimension \footnote{The use of a moving black hole to model cooling
  plasma was first advocated in \cite{zahed}.}. In the above paper, a
holographic renormalization procedure \cite{2} using Fefferman-Graham
coordinates \cite{3} allowed to 
construct the gravity duals of a continuous 1-parametric set of   
4-d stress-energy tensors of the gauge theory containing among
others the duals of the  
{\it free streaming} and {\it perfect fluid} cases. Interestingly, the corresponding family of geometries was shown to possess singularities except only for the latter case, which happens to be a BH moving away in the fifth dimension. This correspondence was shown to be valid at asymptotic proper-times, independently of initial conditions provided boost-invariance is preserved. This gives an AdS/CFT physical criterion for the emergence of perfect fluid behaviour at large proper times.

Starting from this correspondence, it is interesting to study the stability properties of this BH system, since it can bring a physical insight on the typical
relaxation proper-times of the  relativistic thermal and perfect fluid, which does not appear reachable from a direct strong coupling computation in the  ${\cal N}=4$ SYM field
theory. As was done in the static BH case \cite{4} in order to calculate the decay time of an excitation of the system, one computes the  quasinormal modes (QNM) of the Einstein equations linearized around the background  geometry. In particular the calculation of the QNM's corresponding to a scalar excitation  canonically coupled to the metric in the gravitational dual configuration  \cite{4,5,6,6bis} may give an  evaluation of the thermalization time of the dual gauge field-theoretic strongly-coupled system after a small deviation  from equilibrium.

In the present paper, our aim is to extend the analysis from the static case to the black hole moving off in the fifth dimension, deriving the equation and making the
evaluation of the corresponding QNM's in order to 
estimate the thermalization decay proper-time of the relativistically expanding perfect fluid
in the ${\cal N}=4$ SYM field theory. 

The plan of our study is the following. In section {\bf 2}, we shall (re)derive the QNM equation for the static BH using now the Fefferman-Graham coordinates. In 
section {\bf 3} we derive the QNM equation for the gravitational dual of the expanding perfect fluid and give its solutions both in the minimally coupled scalar case and for transverse tensor perturbations. In section {\bf 4}, we discuss the features of our results and their possible physical relevance for the thermalization puzzle of  the QCD quark-gluon plasma in heavy-ion collisions. We conclude and give an outlook.  

\subsection{Quasinormal modes of a static Black Hole in
  Fefferman-Graham coordinates} 

Quasinormal modes (QNM's) formally define the response of a black hole
state to small  
perturbations, 
for instance due to a scalar-field excitation canonically coupled 
to the metric, with incoming (absorbing) boundary condition at the
BH horizon. 
The main observation is that the resulting frequencies become complex
and hence the perturbations are expected to die out
exponentially\footnote{Indeed, for  
cases like Schwarschild background in flat space, there is a power-law
decrease \cite{power}, but it is  
exponential for a Schwarschild background in AdS space. The case of
hydrodynamical long-time  
contributions like  shear viscosity will be discussed later on.}.

Quasinormal
frequencies have been calculated for numerous examples of static black holes
in various numbers of dimensions, in particular for the static planar
AdS black hole which is the dual geometry to $\nn=4$ SYM theory at
nonzero temperature. In this section we shall rederive the known
results using   
the  Fefferman-Graham  coordinates which are suitable for the extension to the
relevant non-static case.  
As already mentioned \cite{1}, the  expanding geometry has  a 
simple form when directly written in these coordinates.

In Fefferman-Graham coordinates the metric of a 5-d planar BH has the form:
\eq
ds^2=\f{1}{z^2} \left[ -\f{\left(1- \f{z^4}{z_0^4}\right)^2}{%
\left(1+ \f{z^4}{z_0^4}\right)} dt^2 + \left(1+ \f{z^4}{z_0^4}\right)
d\vec{x}^2 +dz^2 \right]\ , 
\label{metric}\eqx
where $(t,\vec{x})$ are the boundary coordinates, $z$ is the fifth one, and $z_0$ is the location of the horizon in the bulk.

Quasinormal modes for a scalar perturbation of a black hole are
obtained by solving the wave equation for a massless scalar field
\eq
\Dl \phi \equiv \f{1}{\sqrt{-g}} \partial_i \left(\sqrt{-g} g^{ij}
\partial_j \phi \right)=0
\label{scalar}\eqx
where $g^{ij}$ is the metric tensor and $g$ its determinant in the
background geometry  
\eqref{metric}, assuming purely incoming
boundary conditions at the horizon $z=z_0$ and Dirichlet conditions at the boundary (see e.g. \cite{6bis}).

Inserting eq. \eqref{metric} into \eqref{scalar}, the equation for a
scalar field (with zero  
transverse momentum) has
then the form: 
\eq
-\f{1}{z^3} \f{\left(1- \f{z^4}{z_0^4}\right)^2}{%
  \left(1+ \f{z^4}{z_0^4}\right)} \partial_t^2 \phi(t,z) +\partial_z \left(
\f{1}{z^3} \left(1-\f{z^8}{z_0^8} \right) \partial_z \phi(t,z) \right)\ .
\label{form}\eqx
A separation of variables
\eq \phi(t,z)=e^{i\om t} \phi(z)
\label{wave}
\eqx 
leads to the
ordinary differential equation 
\eq
\partial_z \left(
\f{1}{z^3} \left(1-\f{z^8}{z_0^8} \right) \partial_z \phi(z) \right)
+\om^2 \f{1}{z^3} \f{\left(1- \f{z^4}{z_0^4}\right)^2}{%
  \left(1+ \f{z^4}{z_0^4}\right)}\ \phi(z)=0\ .
\label{flat}\eqx
 Note that, by a change of variable
\eq
z/z_0 = \tanh^{1/4}(4z^*)\ ,
\label{tanh}
\eqx
the equation takes the canonical form
\eq
\partial^2_{z^*} \phi(z^*) 
+\ \f{\sqrt{8}\ e^{12z^*}}{\sinh^{3/2}(8z^*)}\om^2  \phi(z^*)=0
\label{z^*}\eqx
from which one can determine the quasinormal frequencies $\om.$

In fact, by a further change of variable 
\eq
\tilde z = \f{\left(1-(z/z_0)^2\right)^2}{1+(z/z_0)^4} = 1-2 \f 
{\tanh^{1/2}(4z^*)}{1+\tanh(4z^*)}\ ,
\label{tildeh}
\eqx
one shows that this equation can be put in the form of the Heun equation
\eq
\phi''+\f{1-{\tilde z}^2}{{\tilde z} (1-{\tilde z})(2-{\tilde z})}\
\phi' + \f{(\om_{static}/\pi T)^2}{4{\tilde z} (1-{\tilde z})(2-{\tilde z})}\ \phi =0
\label{heun}\eqx
which was 
obtained \cite{5} starting with the conventional BH metric. The QNM
solution  of \eqref{heun}  
dominant at large-time (i.e. with smallest imaginary part)
is found to be  ${\om_{static}}/{\pi T}\sim 3.1194-2.74667\ i$ \cite{5}.

\subsection{Quasinormal modes for the boost-invariant Black Hole geometry}

In this section we derive the equation for a scalar field in the
background corresponding to the asymptotic expanding perfect fluid
geometry. 
The geometry has the form \cite{1,14}:
\eq
ds^2=\f{1}{z^2} \left[ -\f{\left(1- v^4\right)^2}{%
\left(1+ v^4\right)} d\tau^2 + \left(1+ v^4\right)
(\tau^2 dy^2+dx_i^2) +dz^2 \right] 
\label{evolve}\eqx
where $x_{i=1,2}$ are the transverse coordinates, while the proper-time $\tau$ and rapidity $y$ are related to the longitudinal coordinates through $t=\tau \cosh y$ and $x_3=\tau \sinh y$. $v$ is a scaling variable
\eq
\label{e.defv}
v=\f{z}{\tau^{\f{1}{3}}}\ .
\eqx

To this background metric let us couple a scalar field which depends
only on the proper-time 
$\tau,$ and on $z$.
The equation (\ref{scalar}) takes the form:
\eq
-\partial_\tau \left( \f{\tau}{z^3} \f{(1-v^4)^2}{1+v^4}
\partial_\tau \phi \right)  + \partial_z \left( \f{\tau}{z^3} (1-v^8)
\partial_z \phi \right)=0\ .
\eqx
Since the metric \eqref{evolve} has a simple dependence on the
variable $v$ let us 
rewrite the above equation in terms of $\tau$ and $v$. The matrix of
differentials is 
\eq
\label{e.dertau}
\partial_z \to \tau^{-\f{1}{3}} \partial_v \ \ ;\ \ 
\partial_\tau \to \partial_\tau -\f{1}{3}\tau^{-\f{4}{3}} \partial_v\ .
\eqx
Since the perfect fluid geometry is valid in our problem only
at large  proper-time, in the scaling limit $v=const$ and $\tau \to \infty$ we
may consistently neglect the non-diagonal contribution in (\ref{e.dertau}). 
Thus, the QNM calculation preserves the specific  scaling property  in $
{\tau}/{z^3} = 1/v^3$ of the perfect fluid solution of the AdS/CFT
correspondence. 
The resulting equation takes the form:
\eq
-\f{1}{v^3} \f{(1-v^4)^2}{1+v^4} \partial_\tau^2 \phi(\tau,v)
+\tau^{-\f{2}{3}}\ \partial_v \left(\f{1}{v^3} (1-v^8) \partial_v
\phi(\tau,v) \right)=0
\eqx 
which has some  similarity with the one for the
static black hole, once substituting  variables
$(t,z) \to (\tau,v),$ see \eqref{form}.
The noticeable  difference is the additional $\tau^{-2/3}$ factor,
which leads to a nontrivial scaling in the (proper)time dependence. 

Performing a separation of variables $\phi(\tau,v)=f(\tau) \phi(v)$ we
get two decoupled equations. 
An important point is however to notice that the equation for the
$\tau$-dependence  is no 
longer a plane wave as in \eqref{wave} but is
determined by the equation
\eq
\partial_\tau^2 f(\tau)=-\om^2 \tau^{-\f{2}{3}} f(\tau)\ ,
\eqx
whose solutions are linear combinations of the Bessel functions
\eq
\sqrt{\tau} J_{\pm \f{3}{4}} \left( \f{3}{2} \om \tau^{\f{2}{3}} \right)\ .
\eqx
In the large $\tau$ behaviour, the region of validity of the perfect fluid 
geometry
being asymptotic,
the relevant behaviour of the
Bessel functions is
\eq
\label{e.ftau}
f(\tau) \sim \tau^{\f{1}{6}}\ e^{\f{3}{2} i \om \tau^{\f{2}{3}}}\ ,
\eqx
to be compared with the plane waves \eqref{wave} of  the static case.
We will comment on the physical interpretation of \eqref{e.ftau} in the next section.

The resulting $\tau$-independent factorized equation reads
\eq
\partial_v \left(\f{1}{v^3} (1-v^8) \partial_v \phi(v) \right) +\om^2
\f{1}{v^3} \f{(1-v^4)^2}{1+v^4} \phi(v)=0
\eqx
which, strikingly enough, is  formally the same ordinary
differential equation for the 
$\tau$-independent part as 
eq.\eqref{flat} for
the static black hole
but with $v$ playing the role of $z$. Hence we get the same  
profile functions $\phi$ of
the quasinormal modes and their eigenfrequencies $\omega$. However the variables in the perfect-fluid case $v$ and $\tau$ are {\em different} from the $z$ and $t$ variables relevant for the static black hole. Indeed, restoring the $z$ and $\tau$ dependence of the evolving solution $\phi(v)f(\tau)$ gives rise to a rather intricate spacetime dependence.

It is also interesting to derive the quasinormal modes for
perturbations of the metric which on the gauge theory side correspond
to perturbations of the energy-momentum tensor. Of particular interest
are perturbations of the component $T_{x_1x_2}$. Physically their
exponential decay can be interpreted as transverse isotropization of
the asymptotic hydrodynamic expansion of the plasma. On the gravity
side they are interesting since, as shown in \cite{7} for general static cases, their equation of
motion is identical to the scalar one. In particular let us consider
the metric component $g_{x_1x_2}$ and form the quantity
\eq
g^{x_1}_{x_2} \equiv z^2 e^{-c(v,z)} g_{x_1x_2} \ ;\quad  \quad\quad 
e^{c(v,z)} \equiv 1+v^4
\eqx
where $c(v,z)$ is the solution \cite{1} found for the transverse component of the boost-invariant perfect fluid metric (see eq. (\ref{evolve})).

We have shown that in the case of the moving black hole, in the large $\tau$ limit, the quantity $g^{x_1}_{x_2}$ also satisfies also the scalar equation of motion.
\eq
\partial_v \left(\f{1}{v^3} (1-v^8) \partial_v g^{x_1}_{x_2}(v) \right) +\om^2
\f{1}{v^3} \f{(1-v^4)^2}{1+v^4} g^{x_1}_{x_2}(v)=0
\eqx
where the proper-time dependence has been separated out in the same manner
as for the scalar.
Consequently this mode of the metric has the same set of quasinormal
frequencies.

\subsection{Conclusion, discussion and outlook}

Let us summarize our results. We consider the problem of computing the
response proper-time of an expanding 
relativistic and boost-invariant ${\cal N}=4$ SYM
perfect fluid, after a scalar  
excitation off thermodynamic equilibrium. In the dual gravitational
geometry, which corresponds  
to a black-hole  moving away in the fifth dimension, the characteristic
proper-time  is related  
to the frequency $\om_c$ with the smallest imaginary part among quasi-normal modes. 

We derived the corresponding equations in the scalar and the transverse tensor channel.
The resulting equations factorize into two decoupled equations defining respectively  
the explicit proper-time and scaling variable dependence of the modes. 
As in the static case, we find that the quasinormal frequencies have imaginary parts which correspond to exponential decay of perturbations towards the (expanding) equilibrium state. While formally the equation defining the frequencies has the same functional form as in the static case, they correspond to quite different variables namely the proper time $\tau$ and the scaling variable $v=z/\tau^{1/3}$.

The profile of the solution is given as a function of  
the scaling variable $v$, $\phi(v)$, which obeys the same equations as $\phi(z)$ in the static case. By contrast with the
plane waves of the static  
case however, the frequencies are related to Bessel functions
$\sqrt{\tau} J_{\pm \f{3}{4}} \left(  
\f{3}{2} \om \tau^{\f{2}{3}} \right) \to \tau^{\f{1}{6}}\ e^{\f{3}{2} i \om \tau^{\f{2}{3}}},$ hence leading to a non-trivial
scaling in  proper-time. 

These results can be interpreted as follows. The temperature for the expanding perfect fluid is known to behave like $T \sim \tau^{-1/3}$ \cite{bj}, which is consistent through the AdS/CFT correspondence with the evolving black hole solution of \cite{1}.
In order to understand the nontrivial scaling in proper time (\ref{e.ftau}), we note that the {\em static} quasinormal freqencies are proportional to the temperature i.e. $\om_{static}(T)=\al \cdot T$, where $\al$ is a constant.
In the expanding case we can consider an adiabatic approximation assuming that locally the temperature is fixed. Hence a plane wave dependence
\eq
e^{i\,\om_{static}(T) \cdot \tau} = e^{i\al T \tau} \sim e^{i \al \tau^{\f{2}{3}}}
\eqx
will give the scaling in $\tau^{2/3}$ seen in the overall solution. This adiabatic approximation deserves further study. 

A similar discussion is valid for transverse perturbations of the metric. It would be interesting to extend the analysis to generic metric perturbations.

On a theoretical ground, a comment is in order about the well-known
calculation of the viscosity 
to entropy ratio \cite{8} in 
the {\it 
static} BH configuration. Quasinormal modes  and viscosity
calculation  correspond both to   
poles of
specific retarded Green functions for a scalar (for QNM's) and metric
(for viscosity)  
deformation using Fourier transforms in the time  
variable 
$t.$ This is the case for the static case, due to the plane-wave
solution \eqref{wave}. 

The difference between the scalar QNM's and viscosity (shear channel) is that the corresponding
poles do not go to zero at  
small 
transverse momentum contrary to the viscosity case \cite{6bis}. This is expected,
since viscosity is a  
hydrodynamic 
excitation surviving  at large time scales, while normal QNM's lead to an 
exponential 
fall-off, at least within  an AdS space.
 This is the origin of
the finite ``thermalization response-time'' obtained for the scalar equation.

Now turning to the expanding geometry in proper-time, viscosity remains
an interesting issue for  future 
work, since it is not clear whether one can use the method of
Fourier-transformed retarded Green  
functions to 
evaluate it, since the Fourier transforms seem not to be directly relevant for
the proper-time dependence. An adiabatic approximation could help with this problem.  
We postpone this analysis for future work.

In a phenomenological perspective, it is instructive to make a
parallel between our calculation in  
the framework of the ${\cal N}=4$ SYM fluid, and the
problems discussed about the  
thermalization proper-time of a QCD plasma formed in a
heavy-ion collision at high  
energy (see \cite{review} and references therein). Indeed, there are some indication that this typical
proper-time is rather short, which is  
difficult to explain in terms of initial conditions 
dominated by a weak-coupling 
state, and thus with high viscosity and {\em a-priori} long thermalization time. 

In \cite{1}, we have shown that the gauge/gravity
correspondence was selecting the  
perfect fluid solution at large proper-times. Since it seems that the
model of a QCD perfect fluid  
could be favored by the   phenomenological analyses, it is perhaps not
completely unrealistic to  
look for some physical lessons of our present results. The main point
is the rather strong  
stability of the asymptotic solution, since the response-time to a
scalar excitation is short in terms of proper time. Combining eq. \eqref{e.ftau} with the numerical value of the dominant QNM leads to a proper-time damping of the form
\eq
\exp \left( -\f{3}{2} \cdot 2.7466 \cdot \tau^{\f{2}{3}} \right)
\eqx 
in the units where the horizon is fixed at $z_0=\tau^{-1/3}$.

If these estimates would
be also approximately valid  
for QCD, at least in the hypothesis of a deconfined phase for which
the supersymmetric degrees of  
freedom would not play a major r\^ole, this can lead to a new
point-of-view on the thermalization  
problem. Indeed, the stable, strongly interacting state represented by
the perfect gauge fluid,  
could act as an ``attractor'' during the proper-time evolution of a
collision with QCD plasma  
formation, such as for heavy-ion collisions. This would give a typical
non-perturbative mechanism,  
{\em a priori} complementary to those related to the initial perturbative
conditions. It would be   
interesting to see how one could merge the two ends of the evolution,
the perturbative with the  
non-perturbative ones.

\bigskip

\noindent{}{\bf Acknowledgments.} We would like to thank Cyrille Marquet and Gregory Soyez for careful reading of the manuscript and fruitful remarks. RP would like to thank the Institute of Physics, Jagellonian University, Krakow where this work was initiated and RJ thanks the Service de Physique Th\'eorique Saclay where it was finished. RJ was
supported in part by Polish Ministry of Science and Information
Society Technologies grants 2P03B08225 (2003-2006), 1P03B02427
(2004-2007) and 1P03B04029 (2005-2008).


\begin{thebibliography}{99}
\bibitem{14}
For simplicity, we fix here units of $\tau$ such that the horizon is at $z_0\equiv\tau^{-1/3}$.
\end{thebibliography}
\end{document}